\documentclass[11pt]{article}
\usepackage{graphicx}
\usepackage{cs12,html}

\begin{document}
 
\title{The Spectra and Classification of L and T Dwarfs}
 
\author{Adam J.\ Burgasser\altaffilmark{1}}
\altaffiltext{1}{California Institute of Technology; diver@its.caltech.edu}

\index{L dwarfs}
\index{T dwarfs}
\index{Brown dwarfs}
\index{2MASS}
\index{SDSS}
\index{*Gliese 570D}
\index{*2MASS 0559-1404}
\index{*2MASS 0937+2931}
\index{Classification}
\index{Near-infrared spectroscopy}
\index{Optical spectroscopy}

\begin{abstract}

In this contribution, I discuss some of 
the salient features of L and T dwarf spectra
in both the red optical (6300--10000 {\AA}) and near-infrared (1--2.5 $\micron$)
wavelength regimes.  Important absorption bands and lines, including
those of H$_2$O, CH$_4$, FeH, CrH, CO, and the alkali metal lines,
are identified, and
the development of recent classification schemes based on these features are
discussed.  I briefly point out future work required in these wavelength
regions and in the mid-infrared that are required to garner a complete 
picture of these low-luminosity dwarfs.
\end{abstract}

\section{Introduction}

The advent of sensitive near-infrared detectors, and the implementation
of large-scale imaging surveys (e.g., 2MASS, SDSS, and DENIS)
at wavelengths outside of the classical optical regime,
have resulted in a major revolution in cool star research.  
In the past decade, we have witnessed
an avalanche of ultra-cool star and brown dwarf discoveries in the field,
in young stellar clusters, and as companions to nearby stars (see Basri 2000
for an excellent review), pushing to temperatures below 1000 K. 
These discoveries have required the definition
of {\em two} new spectral classes, L dwarfs and T dwarfs, the first major
additions to the widely-accepted MK system in over half a century (see
the contribution by S.\ Hawley in these proceedings for more information).  
Indeed,
the initially unique companion objects GD 165B (Becklin \& Zuckerman 1989) 
and Gliese 229B (Nakajima et al.\ 1995) 
are now understood to be the prototypes to these very late
classes of stars and brown dwarfs.  ``Avalanche'' is an appropriate term
here; despite their recent identification, hundreds of L dwarfs and
at least 30 T dwarfs are now known.

In this contribution, I review some of the spectral properties 
of L and T dwarfs in two well-studied wavelength regimes: first, the 
red optical, typically covering
6300--10000 {\AA}, is discussed
in $\S$2; second, the near-infrared, conventionally 1--2.5 $\micron$, 
is discussed in $\S$3.  In $\S$4, I address future work necessary
in both these spectral regions and beyond that will result in
a more complete understanding of these cool objects.

\section{Red Optical Spectra}

The red optical has been an 
important region of study for the coolest stars, being the brightest
spectral region for these objects accessible by CCD detectors.  
The first spectrum of an L dwarf, GD 165B,
was obtained in the red optical by Kirkpatrick et al.\ (1993), and
shown to lack the strong
TiO bands that characterize M dwarfs.  With the
discovery of additional ``post-M'' objects (see 
Basri 2000), it became clear
that a new spectral class, the L dwarfs, had been identified.
For the T dwarfs, initial work in this regime was done by Oppenheimer
et al.\ (1998) in their analysis of Gliese 229B.  These authors
identified a number of
distinct features, including a steep red slope extending from
8600--10000 {\AA} that initially defied adequate explanation.

\subsection{L Dwarfs}

Figure 1 shows the spectral sequence of late-M and L dwarfs in the red optical
from Kirkpatrick et al.\ (1999); see Fig.\ 3
in the contribution by S.\ Hawley for the identification of the 
features present in these spectra.
As originally
observed in GD 165B, the hallmark TiO and VO bands weaken in the L dwarfs,
resulting in a fairly
transparent atmosphere in which hydrides and alkali lines become 
prominent.  FeH (8692, 9896 {\AA}), CrH (8611, 9969 {\AA}),
and CaH (6750 {\AA}) are all present, strengthening in the early-type
L dwarfs and weakening in the later types.  H$_2$O also produces an important
feature at 9250 {\AA}, which strengthens toward later spectral types and is
important in the T dwarfs as well.

\begin{figure}[h]
\centering
\includegraphics*[width=7cm]{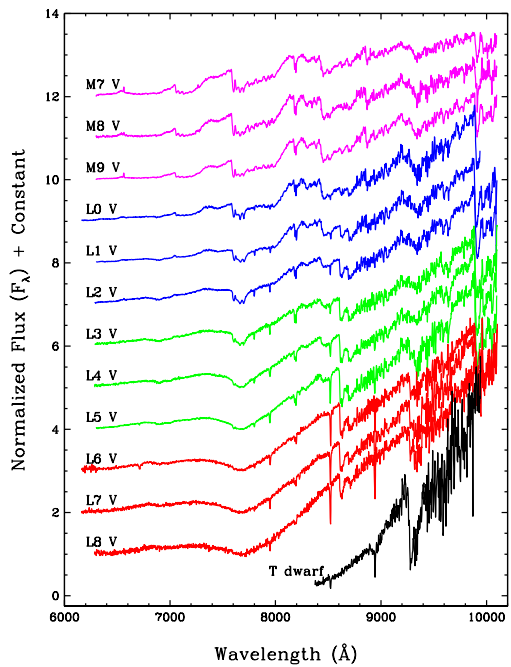}
\caption{Red optical spectra of late-M and L dwarfs, adapted
from Fig.\ 6 in Kirkpatrick et al.\ (1999).  
Subtype designations are from that reference. 
See Fig.\ 3
in the contribution by S.\ Hawley for identification of features.}
\end{figure}

The depletion of molecular 
species, likely caused by the formation and rainout of dust (Tsuji, Ohnaka,
\& Aoki 1996; Ackerman \& Marley 2001), allows the
less refractory
alkali doublet lines to come into prominence, including Na I (5890/5896
and 8183/8195 {\AA}), K I (7665/7699 {\AA}), Cs I (8521/8943 {\AA}), and
Rb I (7800/7948 {\AA}).  The large
column abundances of both Sodium and Potassium in these clear atmospheres
enable their resonance doublet features
to become quite pronounced via pressure-broadening.
This can be seen in the case of K I by the wide
trough centered at
7700 {\AA} in the latest L dwarfs, growing from the
sharper features present in the late M dwarfs and early L dwarfs.
The Na I 5890/5896 {\AA} D lines
produce a similar feature at shorter wavelengths (Reid et al.\ 2000).  
The Li I line at
6708 {\AA} is also present in L dwarfs 
with masses below 0.06M$_\odot$; Rebolo et al.\ (1992) first pointed out
the utility of this feature as an indicator for substellarity.  Kirkpatrick
et al.\ (1999) estimate that over 1/3 of the field L dwarfs 
identified by 2MASS are brown dwarfs,
based on the presence of this line in their spectra. 
The other line feature frequently seen in these
spectra is H$\alpha$ emission at 6563 {\AA}.  Contributions by S.\ Hawley
and S.\ Mohanty provide
further discussion on H$\alpha$ emission in L dwarfs.

The seemingly smooth variation of these features in the observed L dwarf
population has resulted in two competing classification schemes: that of 
Kirkpatrick et al.\ (1999), which spans subtypes L0 V to L8 V; and that
of Mart{\'{\i}}n et al.\ (1999), which spans subtypes L0 V to L6 V.  
Utilizing both band and color indices to classify subtypes, these
systems agree fairly well for the early and mid-L dwarfs, but diverge 
beyond type L4.  Hopefully the current treasury of L dwarf spectra will allow
convergence of these two schemes in the near future.

\subsection{T Dwarfs}

The last spectrum in Figure 1 is that of the prototype T dwarf, Gliese 229B,
from
Oppenheimer et al.\ (1998).  Its spectrum is quite different that those of the
latest L dwarfs, due to deeper H$_2$O absorption, absence of hydrides,
and steeper spectral slope.  However, it was unclear at the time of
its discovery if
the photosphere of Gliese 229B is influenced by its primary, 
which could naturally lead to an unusually red spectrum (Griffith, 
Yelle, \& Marley 1998).

\begin{figure}
\hspace{0.5cm}
\includegraphics*[width=6cm]{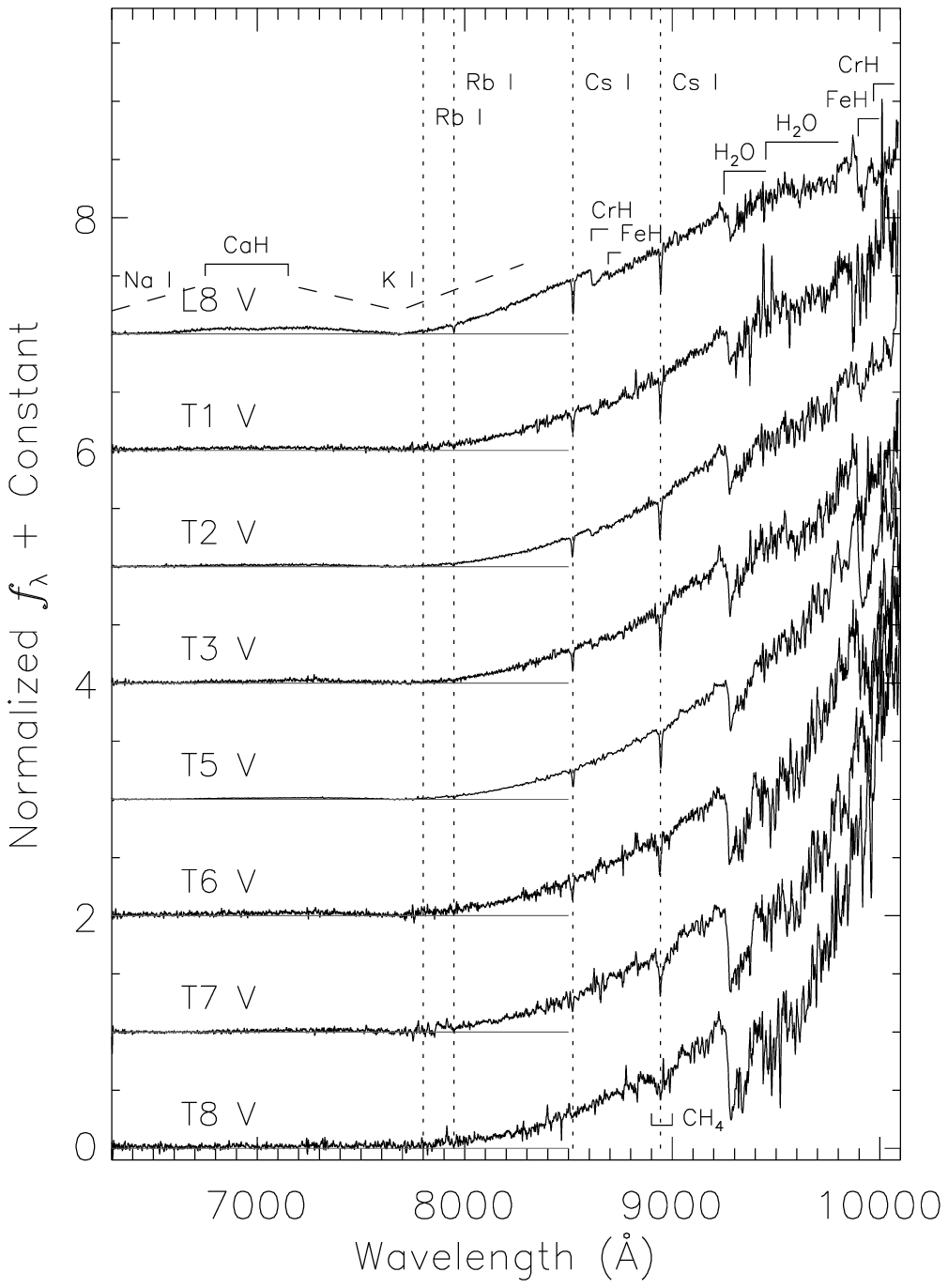}
\vbox{\includegraphics*[width=6cm]
{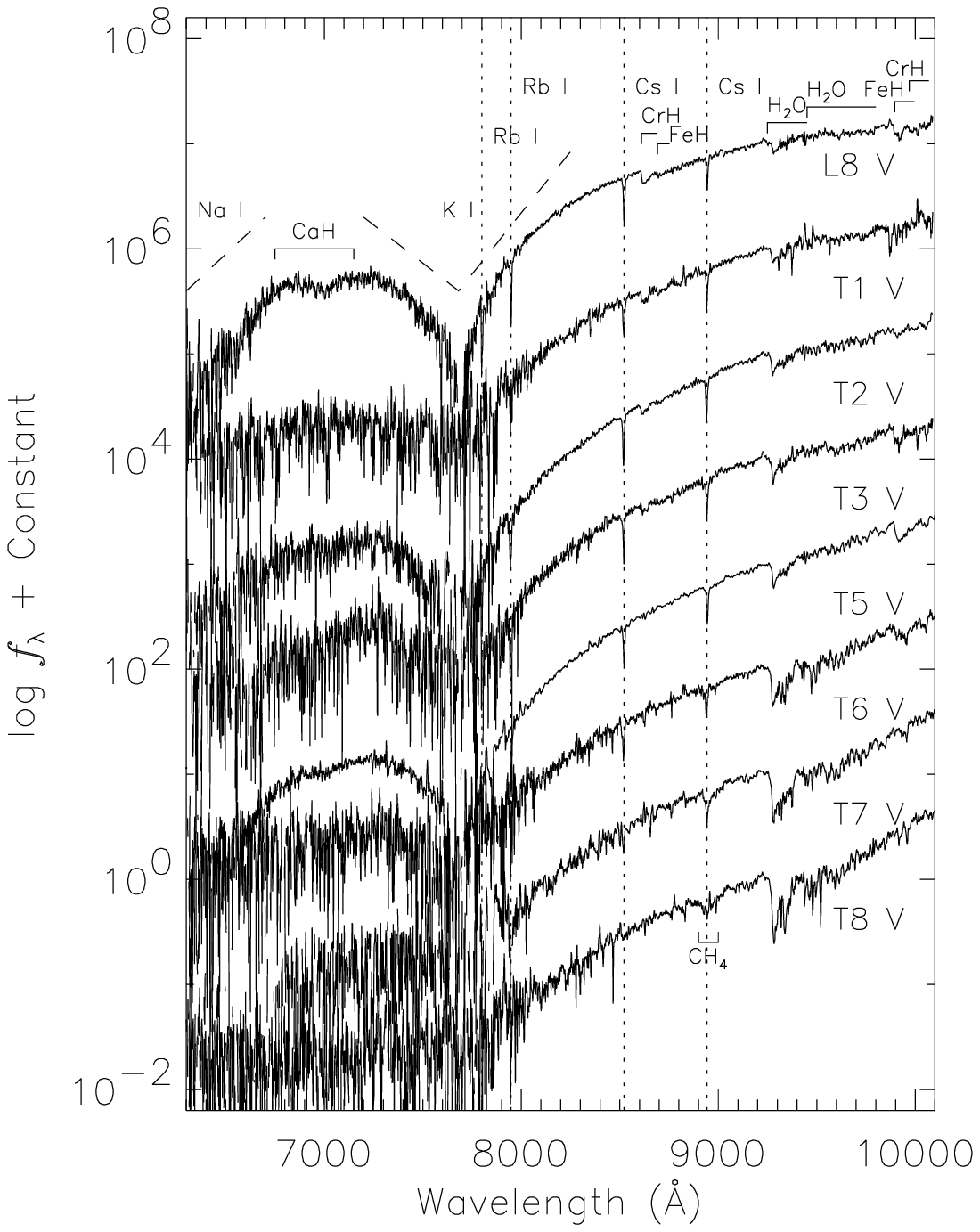}}
\caption{Red optical spectra of T dwarfs 
from Burgasser (2001).  The left panel shows spectra on a linear
scale, the right on a logarithmic scale.
Molecular absorption bands of CaH, FeH, CrH, H$_2$O, and CH$_4$ are
indicated, as are alkali lines.  The heavily broadened Na I and K I 
features are indicated by dashed lines.  Subtypes are from 
Burgasser et al.\ (2001).}
\end{figure}

Figure 2 shows red optical spectra of a series of T dwarfs and one
L8 dwarf from Burgasser (2001); subtypes are those defined in Burgasser
et al.\ (2001) (see $\S$3.2).  All of these T dwarfs are 
field objects and show the
same steep spectral slope in the red as Gliese 229B. Liebert et al.\ (2000)
have shown that the increasingly broadened
K I doublet seen in the L dwarfs
is generally responsible for this slope and the corresponding
red optical/near-infrared colors of T dwarfs, as is obvious in the spectra
in the right panel of Figure 2.  Na I D lines
suppress flux shortward of
7000 {\AA}, an extension of the behavior seen in the latest L dwarfs.
There are also a number of unbroadened atomic features present in these
spectra, including
Cs I at 8521/8943 {\AA} and the much weaker
Rb I doublet at 7800/7948 {\AA}.
The former lines appear to be
strong throughout the T sequence, although they fade rapidly in the latest
subtypes.
The latter are only seen in the highest signal-to-noise spectra
and are generally weak, buried in the broadened K I feature.
Note that Li I at 6708 {\AA}
has not yet been detected in any of the T dwarfs.

The most prominent molecular features in the red optical spectra of T dwarfs
are the H$_2$O bands starting at
9250 and 9450 {\AA}, respectively.  These bands
strengthen considerably
throughout the T sequence, and provide powerful diagnostics for classification.
An FeH band at 9896 {\AA}
is also seen in some of the spectra of Figure 2,
appearing to strengthen somewhat between the latest L dwarfs and
the mid-T dwarfs, then 
fading again toward the latest T subtypes.
Its higher-order
counterpart at 8692 {\AA}
is seen only in the earliest T dwarfs, which is also the case for
CrH at 8611 {\AA}.  The 9969 {\AA} band of CrH is weak or absent
in all of the T dwarfs.
A weak signature of CaH around
6750--7150 {\AA}, seen in M and L
dwarfs, also appears to be present but weak in the brightest early T dwarfs.
Finally, 
CH$_4$, which is ubiquitous in the near-infrared spectra of T dwarfs
(see $\S$3.2), produces a weak band centered at 8950 {\AA} in the latest
subtypes.  The presence of this feature is made ambiguous
by its coincidence with Cs I at 8943 {\AA}.

Finally, steady
H$\alpha$ emission has been detected in the spectrum of 2MASS 1237+6526
(Burgasser et al.\ 2000b), and a much
weaker line has been seen in the spectrum of 2MASS 1254-0122 (Kirkpatrick 
et al.\ 2002).  
None of the other T dwarf spectra show emission features.
We refer the reader
to the contribution of S.\ Hawley in these proceedings
for further discussion of H$\alpha$ emission in T dwarfs.

\section{Near-infrared Spectra}

The near-infrared is an useful region for the study of L and T dwarfs, as
the spectral energy distributions of both classes peak at roughly 1 $\micron$.
The first spectrum of an L dwarf, GD 165B,
was obtained by Jones et al.\ (1994), who 
noted the importance of H$_2$O and CO
absorption in these cool dwarfs, as well as their apparent reddening
as compared to warmer M dwarfs (cause by thermal dust emission).
Nearly all of the spectroscopic work on T dwarfs has been done in the
near-infrared due to sensitivity considerations\footnote{For Gliese 229B,
R--J $\sim$ 9 (Golimowski et al.\ 1998).}.
Indeed, it is the difference in the near-infrared spectra of
L dwarfs and Gliese 229B due to CH$_4$ absorption
that prompted Kirkpatrick et al.\ (1999) to designate
a new class for these cool brown dwarfs.  

\subsection{L Dwarfs}

Figure 3 shows the near-infrared spectra of three late M dwarfs and three
L dwarfs
from Leggett et al.\ (2001).  The deepening of the H$_2$O bands
at 1.15, 1.4, and 1.85 $\micron$ for these objects
is readily apparent in the data.
Not surprisingly, the H$_2$O bands split the
near-infrared spectra of L dwarfs into peaks centered at
the three telluric windows, around 1.2, 1.65, and 2.15 $\micron$; 
the L dwarf H$_2$O bands
are significantly broader than telluric absorptions, 
however, due to the higher temperatures of  
L dwarf atmospheres.  The relative brightness of these
flux peaks (increasing importance of K-band)
reflects the reddening of L dwarfs in near-infrared colors.

\begin{figure}
\centering
\includegraphics*[angle=270, width=10cm]{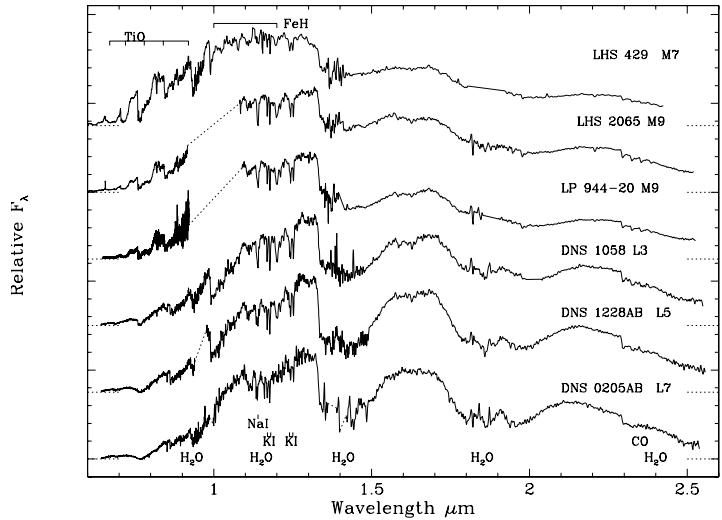}
\caption{Near-infrared spectra of three late M dwarfs and three L dwarfs
from Leggett et al.\ (2001).  Major absorption bands of TiO, FeH, H$_2$O,
and CO are indicated, as are atomic lines of Na I and K I.}
\end{figure}

CO is a key absorber at K-band; the bandheads beyond 2.3
$\micron$ are clearly seen in the M9 dwarfs, with the some slight
strengthening in the L dwarfs.  The latest-type objects also show absorption
beyond the 2.1 $\micron$ peak; this was originally attributed to 
CH$_4$ by Delfosse et al.\ (1997),
but later identified as collision-induced absorption (CIA) of H$_2$
by Tokunaga \& Kobayashi (1999).  CIA H$_2$ is an 
important contributor to the total opacity in the cold, high-gravity 
atmospheres of both L and T dwarfs.  Recently, Geballe et al.\ (2001) have found
indications of weak CH$_4$ absorption at K-band in their latest L9 dwarfs
in addition to the CIA H$_2$ opacity.

There are finer features present at J-band in the L dwarfs.
In particular, K I doublet lines at
1.169/1.177 and 1.243/1.252 $\micron$ are quite strong in the L dwarfs,
weakening somewhat for the latest subtypes (McLean et al.\ 2000).  
These are higher-order
lines than the 7700 {\AA} resonance doublet, and hence do not broaden
as strongly.  Na I lines at 1.138/1.141 $\micron$ show similar behavior.
Additional alkali lines of Rb I and Cs I are likely present but
buried in the broad 1.4 $\micron$ H$_2$O band.  FeH features at
1.194, 1.210, and 1.237 $\micron$ are present and strong in the
early subtypes, eventually
fading along with the 9896 {\AA} band (McLean et al.\ 2000).

In analogy to the red optical, the evolution of features in the near-infrared
has allowed the derivation of classification schemes for L dwarfs in this
wavelength regime.  Reid et al.\ (2001) and Testi et al.\ (2000)
have made use of 
H$_2$O and color indices to derive spectral classification schemes
tied to the Kirkpatrick et al.\ (1999) optical sequence.
Geballe et al.\ (2001) have defined an
independent near-infrared classification scheme that extends to subtype L9,
consistent with both the Kirkpatrick et al.\ (1999) and 
Mart{\'{\i}}n et al.\ (1999) optical classifications at early types,
but diverging somewhat for the latest subclasses.  See the contribution by
T.\ Geballe in these proceedings for further discussion.  
It is hoped that ongoing analysis
will ultimately produce a consistent
system that matches red optical and near-infrared subclasses.

\subsection{T dwarfs}

Figure 4 shows low-resolution spectra of T dwarfs and two L dwarfs
in the 1--2.5 $\micron$
region, with important molecular and atomic features indicated (higher
resolution spectra are given in the contribution by T.\ Geballe).
These data show that both H$_2$O and CH$_4$
absorption bands are the dominant features shaping the spectra of T dwarfs. 
The addition of CH$_4$
bands (beginning at 1.05, 1.3, 1.6, and 2.2 $\micron$) to the H$_2$O bands
seen in the L dwarfs ultimately confine the
emitted flux of these objects into narrow peaks centered at 
1.08, 1.27, 1.59, and 2.07
$\micron$.  CH$_4$ absorption effectively removes half of the emitted
flux from the H- and K-bands, while the 
gradually developing slope between the
1.27 $\micron$ peak and the 1.4 $\micron$ H$_2$O band is caused by
CH$_4$ at 1.3 $\micron$.  Earlier-type T dwarfs than those shown here
have weaker CH$_4$ absorption, resulting in the dual presence of
CH$_4$ and CO bands at 2.2 and 2.3 $\micron$ 
(Leggett et al.\ 2000).  CIA H$_2$ is also present at
K-band, suppressing flux throughout the 2--2.5
$\micron$ region\footnote{CIA H$_2$ absorption also peaks around 1.2
$\micron$ but is 100 times weaker.}.  Variation in the strength of
H$_2$ absorption amongst the known T dwarf population
has been noted by Burgasser et al.\ (2001) in the peculiar T dwarf
2MASS 0937+2931, possibly indicative of extreme gravity and/or metallicity
differences.  See the contribution by S.\ Leggett in these 
proceedings for further observations of this effect. 

\begin{figure}
\centering
\includegraphics*[angle=90, width=10cm]{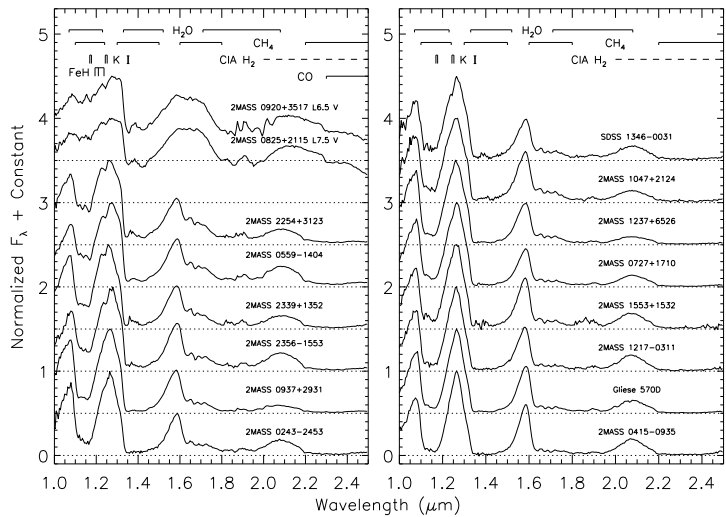}
\caption{Low resolution 
near-infrared spectra of T dwarfs and two late-type L dwarfs
identified by 2MASS.  Major absorption bands of H$_2$O, CH$_4$, CIA H$_2$,
and CO are indicated, along with the location of FeH (in the L dwarfs)
and K I features at J-band (from Burgasser et al.\ 2001).}
\end{figure}

Atomic lines of K I at J-band appear to remain
strong from the L dwarfs through most of the T dwarfs, but
weaken somewhat in the very latest T subtypes.  Indeed, both
Burgasser et al.\ (2001) and Geballe et al.\ (2000a) note an absence
of these lines in the coolest known T dwarf Gliese 570D
(Burgasser et al.\ 2000a). The spectra of Figure 4 are generally too low
to resolve the K I lines, but their trends can be seen in the
evolution of the notch feature around 1.25 $\micron$.  FeH lines
seen in the L dwarfs at J-band are not seen in the T dwarfs.

\begin{figure}
\centering
\includegraphics*[width=8cm]{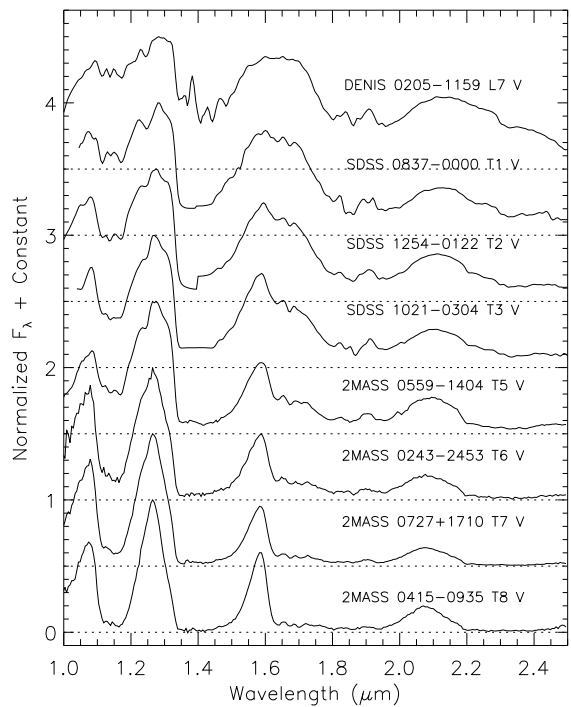}
\caption{Near-infrared spectra of T dwarf standards and the L7 V 
DENIS 0205-1159AB, showing the evolution of spectral features with
subtype (from Burgasser et al.\ 2001). 
See the contribution by T.\ Geballe for further details on
T dwarf classification.}
\end{figure}

T dwarf classification has only recently been addressed by Burgasser et al.\
(2001) and Geballe et al.\ (2001), with both authors using near-infrared
features of H$_2$O, CH$_4$, and color to derive very similar schemes.
We defer discussion of T classification to the contribution of T.\ Geballe,
but show in Figure 5 the spectral standards from Burgasser et al.\ (2001).
These data 
highlight the clear strengthening of band features and weakening of 
the K I lines throughout the sequence. 

\section{Future Work}

Since their first discovery, there have been significant advances made in
the observational study of brown dwarfs.  The focus
of these studies has primarily been in the red optical and near-infrared,
enabling the thorough characterization of brown dwarfs in these wavelength
regimes as described in this article. 
There is a great deal to be learned in this region, however, particularly
in the interpretation of the spectra.  Theoretical modeling of
brown dwarf atmospheres has advanced greatly (see Burrows et al.\ 2001 
and Chabrier \& Baraffe (2000) for
recent reviews), but 
significant work remains in detailing the 
differential effects of temperature, gravity, and 
metallicity on emergent spectra; updating molecular opacities for
adequate modeling; and incorporating self-consistent treatment of 
dust and grain opacities, including cloud formation and rainout
(see the contribution by M.\ Marley in these proceedings).  
The latter problem is particularly critical in 
understanding the transition between the dusty L and relatively
dust-free T subtypes. 

There has been interesting work recently
done outside of the 0.6--2.5 $\micron$
regime for both L and T dwarfs.  Strong CO absorption has been detected in the 
T dwarfs Gliese 229B (Noll, Geballe, \& Marley 1997,
Oppenheimer et al.\ 1998) and
2MASS 0559-1404 (Burgasser 2001), indicative of upwhelling in a dynamic
photosphere, a complex process requiring more advanced theoretical
modeling (Saumon et al.\ 2000).
In addition, the 
fundamental band of CH$_4$ at 3.3 $\micron$
has been detected in mid- to late-L dwarfs, which has possible implications
on the temperature scale of L and T subclasses
(Noll et al.\ 2000).  Further work beyond
5 $\micron$, which can be accomplished via SIRTF and SOFIA, 
could potentially enable
the direct detection of dust species in this objects, as well as strong bands
of NH$_3$, H$_2$S, and alkali chlorides (the repository for atomic alkali
species below $\sim$800 K; Lodders 1999).  The spectra of L and 
T dwarfs clearly have
many secrets left to be revealed.

\acknowledgments

I would like to acknowledge all of my collaborators on the 2MASS Rare Objects
Team: A.\ Burrows, R.\ M.\ Cutri, C.\ C.\ Dahn, J.\ E.\ Gizis, 
J.\ D.\ Kirkpatrick, J.\ Liebert, 
P.\ J.\ Lowrance, D.\ G.\ Monet, I.\ N.\ Reid, and  M.\ F.\ Skrutskie; and my
advisor M.\ E.\ Brown at Caltech.  I recognize the support 
of the Jet Propulsion
Laboratory, California Institute of Technology, which is operated under
contract with the National Aeronautics and Space Administration. 

Online resources for L and T dwarf spectra may be found
at I.\ N.\ Reid's 
\htmladdnormallink{L and M dwarf spectra page}{http://dept.physics.upenn.edu/~inr/ultracool.html},
D.\ Montes'
\htmladdnormallink{L and M dwarf spectra page}{http://www.ucm.es/info/Astrof/fgkmsl/mldwarfs.html},
C.\ Gelino's 
\htmladdnormallink{Brown Dwarf Catalog}{http://ganymede.nmsu.edu/crom/cat.html},
S.\ Leggett's
\htmladdnormallink{Spectra Repository}{ftp://ftp.jach.hawaii.edu/pub/ukirt/skl/},
and my own
\htmladdnormallink{T Dwarf page}{http://www.gps.caltech.edu/~pa/adam/}.



\begin{references}

\newcommand\nat{{Nature}} 

\reference Ackerman, A.\ S., \& Marley, M.\ S.
2001, \apj, 556, in press 

\reference Basri, G. 2000, \araa, 38, 485

\reference Becklin, E.\ E., \& Zuckerman,
B. 1988, Nature, 336, 656 

\reference Burgasser, A.\ J. 2001, Ph.D.\ Thesis, California Institute of Technology

\reference Burgasser, A.\ J., et al. 2001, \apj, 563, in press

\reference Burgasser, A.\ J., et al. 2000a, \apj, 
531, L57

\reference Burgasser, A.\ J., Kirkpatrick, J.\ D.,
Reid, I.\ N., Liebert, J., Gizis, J.\ E., \& Brown, M.\ E. 2000b, \aj, 120,
473

\reference Burrows, A., Hubbard, W.\ B., Lunine, 
J.\ I., \& Liebert, J. 2001, Rev.\ of Modern Physics, in press

\reference Chabrier, G., \& Baraffe, I.
2000, \araa, 38, 337

\reference Delfosse, X., et al. 1997, \aap, 
327, L25

\reference Geballe, T.\ R., et al. 2001b, \apj, 563, in press

\reference Golimowski, D.\ A., Burrows, C.\ S., Kulkarni, S.\ R.,
Oppenheimer, B.\ R., \& Brukardt, R.\ A. 1998, \aj, 115, 2579

\reference Griffith, C.\ A., Yelle, 
R.\ V., \& Marley, M.\ S. 1998, Science, 282, 2063

\reference Jones, H.\ R.\ A., Longmore, A.\ J., 
Jameson, R.\ F., \& Mountain, C.\ M. 1994, \mnras, 267, 413

\reference Kirkpatrick, J.\ D., et al. 2002,
\aj, in prep. 

\reference Kirkpatrick, J.\ D., et al. 1999, 
\apj, 519, 802 

\reference Kirkpatrick, J.\ D., Kelly, D.\ M., 
Rieke, G.\ H., Liebert, J., Allard, F., \& Wehrse, R. 1993, \apj, 402, 643

\reference Leggett, S.\ K., Allard, F.,
Geballe, T., Hauschildt, P.\ H., \& Schweitzer, A. 2001, \apj, 548, 908

\reference Leggett, S.\ K., et al. 2000,
\apj, 536, L35

\reference Liebert, J., Reid, I.\ N., Burrows, A.,
Burgasser, A.\ J., Kirkpatrick, J.\ D., \& Gizis, J.\ E. 2000, \apj, 533,
L155

\reference Lodders, K. 1999, \apj, 519, 793

\reference Mart{\'{\i}}n, 
E.\ L., Basri, G., \& Zapatero Osorio, M.\ R.
1999, \aj, 118, 1005 

\reference McLean, I., et al. 2000, \apj, 533, L45

\reference Nakajima, T., Oppenheimer, B.\ R., 
Kulkarni, S.\ R.,
Golimowski, D.\ A., Matthews, K., \& Durrance, S.\ T. 1995, \nat, 378, 463

\reference Noll, K.\ S., Geballe, T.\ R., Leggett, 
S.\ K., \& Marley, M.\ S. 2000, \apj, 541, L75

\reference Noll, K.\ S., Geballe, T.\ R., \&
Marley, M.\ S. 1997, \apj, 489, L87

\reference Oppenheimer, B.\ R., Kulkarni, S.\ R.,
Matthews, K., van Kerkwijk, M.\ H. 1998, \apj, 502, 932

\reference Rebolo, R., 
Mart{\'{\i}}n, E.\ L., \& Magazzu, A. 1992, \apj, 389, L83

\reference Reid, I.\ N., Burgasser, A.\ J., 
Cruz, K., Kirkpatrick, J.\ D., \& Gizis, J.\ E. 2001, \aj, 121, 1710

\reference Reid, I.\ N., Kirkpatrick, J.\ D.,
Gizis, J.\ E., Dahn, C.\ C., Monet, D.\ G., Williams, R.\ J., 
Liebert, J., \& Burgasser, A.\ J. 2000, \aj, 119, 369 

\reference Saumon, D., Geballe, T.\ R., 
Leggett, S.\ K., Marley, M.\ S., Freedman, R.\ S., Lodders, K.,
Fegley, B., Jr., \& Sengupta, S.\ K. 2000, \apj, 541, 374

\reference Testi, L., et al. 2001, \apj, submitted

\reference Tokunaga, A.\ T., \& 
Kobayashi, N. 1999, \aj, 117, 1010

\reference Tsuji, T., Ohnaka, K., \&
Aoki, W. 1996, A{\&}A, 305, L1


\end{references}
\end{document}